\documentclass[%
 amsmath,amssymb,
 aps,
prl,
]{revtex4-2}

\usepackage{graphicx}
\usepackage{dcolumn}
\usepackage{bm}
\usepackage{natbib}
\usepackage{mathtools}
\usepackage{hyperref}

\begin{document}

\title{Is the BKL map an intrinsic feature of the Quantum Mixmaster Universe?}
\thanks{To the memory of Patrizio Belli}%

\author{Simone Lo Franco}
 \email{simone.lofranco@uniroma1.it}
\affiliation{%
    Department of Physics, Sapienza University of Rome, Piazzale Aldo Moro, 2, 00185 Roma RM
}%

\author{Giovanni Montani}
\email{giovanni.montani@enea.it}
\affiliation{
    ENEA, Fusion and Nuclear Safety Department, C.R. Frascati, Via E. Fermi, 45, Frascati, 00044, Roma, Italy
}%
\affiliation{
    Department of Physics, Sapienza University of Rome, Piazzale Aldo Moro, 2, 00185 Roma RM
}%

\date{\today}

\begin{abstract}
We study the quantum Mixmaster dynamics by constructing the corresponding Wheeler-DeWitt equation as a relativistic quantum theory in a pseudo-Riemannian Mini-superspace. The transition amplitude from a Kasner regime to the next one is built via the standard S-matrix of a relativistic scattering, and mediated by a single wall potential. We provide a coherent and convincing representation of the original Misner's idea that quasi-classical states can survive arbitrarily close to the cosmological singularity. By constructing a complete set of states, we demonstrate that the quasi-classical BKL map is preserved by the resulting transition amplitude to a new state when their mean total momentum is high enough (exactly coinciding with Misner's high occupation numbers). This scheme is naturally iterated to the initial singularity without appreciable shape changes. We also clarify the role played by the state localization, and, in particular, we outline that when the incoming state is spread, the possibility of violating the Kasner map is no longer negligible. Thus, Kasner-like wave packets can approach the initial singularity. 
\end{abstract}

\maketitle

\section{Introduction}
One of the most important results obtained by the Landau School concerns the derivation of the oscillatory regime characterizing the asymptotic dynamics of the homogeneous Bianchi type VIII and IX models towards the cosmological singularity \cite{BKL_1971,Montani:PrimCos,montani_mix_review}. This achievement acquired particular impact on the modern concept of relativistic cosmology because of the generalization of Bianchi VIII and IX chaoticity also to small enough spatial paths of a generic inhomogeneous cosmology \cite{BKL_1982,kirillov_93,Montani_1995,Montani:PrimCos,BL_1992, Barrow:2020rsp}. Immediately after the assessment of these studies, C.W. Misner provided a Hamiltonian formulation of the same homogeneous chaotic Bianchi IX cosmology, and he dubbed the resulting scheme \emph{Mixmaster Universe} \cite{Misner_mixmaster68} (see \cite{cornish_levin_97,montani_imponente_01} for a covariant formulation of chaos). Also, the Mixmaster formulation has been extended to the inhomogeneous sector, as in \cite{kirillov_93,montani_benini_04}, so that a point-like 2D Hamiltonian dynamics is today a basic feature in the very early Universe description. Once B. DeWitt provided the derivation of the canonical formulation for quantizing the gravitational field \cite{DW_QC_1,DW_QC_2,DW_QC_3}, Misner himself provided the first implementation of this theory to the Bianchi IX cosmology \cite{misner_qc1} (see \cite{montani_benini2006inhomogeneous,montani_antonini2019singularity} for an inhomogeneous extension). In this study, Misner treats the cosmological dynamics as that one resembling a pin-point particle in a receding well. He also implemented an adiabatic formulation, according to which the dependence of the Universe wave function on the space volume is weaker with respect to the anisotropy degrees of freedom, being the fundamental physical ingredients. Despite this relevant modelization of the real canonical quantum dynamics, an outstanding result emerged in \cite{misner_qc1}: semiclassical states, described by high values of the anisotropy occupation numbers, can survive arbitrarily close to the initial singularity (being not removed in this context). A significant effort to construct a Hilbert space for the Bianchi cosmologies has been performed in \cite{Wald_1993,Higuchi_1995}, where the pseudo-Riemannian character of the misupermetric \cite{montani_cqg} has been implemented, allowing the identification of the isotropic Misner variable as a time coordinate. In \cite{Giovannetti:2022qje,giovannetti_maione,LoFranco:2024nss}, the standard methodology of a scattering problem in Relativistic Quantum Mechanics, as presented in \cite{Bjorken:Drell}, has been implemented in the Mini-superspace aiming to characterize the emerging Big-Bounce as a quantum transition process \cite{Ashtekar_2010, Barca_2021}. In this work, we use the theoretical background provided in \cite{Wald_1993} to describe the quantum Mixmaster dynamics as a scattering process. However, differently from the analysis of \cite{Giovannetti:2022qje,giovannetti_maione,LoFranco:2024nss}, here we study only transition amplitude between positive frequency solutions. In particular, according to the equilateral triangular symmetry of the Bianchi IX potential, the potential mediating the scattering between inward and outward states is one characterizing a single wall, i.e., a Bianchi II-like spatial curvature. The two main aims here are: 
i) studying the conditions under which the reflections against the potential wall follow the quasi-classical BKL map, so validating Misner's idea; 
ii) determining the existence of non-negligible transition amplitude that, for suitable states, allows the violation of the BKL map, stopping the quantum Mixmaster behavior.

\section{The Bianchi II model as the building block of Mixmaster Universe}\label{BIX_MXMST}
In this section, we present descriptions of the Bianchi IX model and how its dynamics is related to that generated by the Bianchi II Universe. For this classical system description, we follow the Arnowitt-Deser-Misner formulation of GR in terms of the Misner variables as discussed in \cite{Misner:1973prb,Montani:PrimCos}. In the ADM formalism, the action of the Bianchi IX model in the vacuum reads as \cite{Montani:PrimCos}
    \begin{equation}%
        \mathcal{S}_{IX} = \int dt\ \left( p_a \dot{q^a} - N(t) \mathcal{H}_{IX}\right)\,,
        \label{BIX-action}%
    \end{equation}%
    \label{BIX-action-n-ham}%
Here, the generalized coordinates $q^{a}$ are the three exponential scale factors of the three-metric and $p_a$ their conjugated momenta. $\mathcal{H}_{IX}$ is the Super-Hamiltonian, while $N(t)$ is the lapse function fixed by the choice of reference system. To deal with a diagonal kinetic term in momenta, we use the \emph{Misner coordinates} $\{ \alpha,\, \beta_+,\, \beta_- \}$, which are connected to the original ones through a canonical transformation \cite{Misner:1973prb}. The coordinate $\alpha$ is related to the Universe volume while $\beta_\pm$ account for the spatial anisotropies. Hence, the Super-Hamiltonian reads as 
\begin{equation}%
    \mathcal{H}_{IX} = \frac{\chi e^{-3\alpha}}{3(8\pi)^2} \left[ -p^2_\alpha + p^2_+ + p^2_- + u\, U_{IX}(\alpha, \beta_\pm) \right],
    \label{BIX-ham-misner}%
\end{equation}%
where $p_\alpha,\, p_\pm$ are the momenta conjugated to the Misner coordinates, $\chi \equiv8\pi G/c^4$, and $u\equiv 3 (4\pi)^4/\chi^2$. Here, $U_{IX}$ is the potential term generated by the spatial curvature of the model, and it is a positive-definite potential well having the symmetries of an equilateral triangle. The Super-Hamiltonian in Eq. (\ref{BIX-ham-misner}) is analogous to that of a particle moving in the potential $U_{IX}$ with the difference that its kinematical part is not positive-definite; this feature will allow us to treat the WDW equation of this system as a Klein-Gordon one. We now focus on the features of the Bianchi IX Universe near the cosmological singularity. In the regions where $U_{IX}(\beta_\pm)$ is negligible, the Universe is described by the usual Kasner solution for a Bianchi I model, that in the gauge $N=3(8\pi)^2e^{3\alpha}/(2\chi)$ reads as
    \begin{equation}%
        \alpha(t) = -p_\alpha\, t\,, \quad \beta_+(t) = -\frac{p_+}{p_\alpha} \alpha\,, \quad \beta_-(t) = -\frac{p_-}{p_\alpha} \alpha\,,
    \end{equation}%
where $p_\alpha = p_+ = p_- = const$. Denoting with $\beta'_\pm \equiv d\beta_\pm/d\alpha$ the components of the \emph{anisotropy velocity}, the well-known condition on the Kasner exponents is equivalent to $|\bm{\beta}'|^2\equiv|\beta'_+|^2+|\beta'_-|^2 = 1$. As $\alpha \to -\infty$, $U_{IX} \approx U_{II} =\exp{[4\alpha-8\beta_+]}$. Therefore, when $\beta'_+ > 1/2$, the contribution from the potential is large enough to influence the dynamics, and the Universe will collide on the potential wall. Then, the Universe continues evolving in a new Kasner phase until it interacts again with the potential. In this way, the Universe follows an infinite sequence of collisions \cite{Misner:1973prb}. To investigate the early stages of Universe history, we consider the transformation $\alpha \to -\alpha$, as the initial singularity is now placed at $\alpha \to +\infty$. Therefore, for a single collision, the Bianchi IX Universe is described by the super-Hamiltonian 
\begin{equation}
    \mathcal{H}_{II} = \frac{\chi e^{-3\alpha}}{3(8\pi)^2} \left[ -p^2_\alpha + p^2_+ + p^2_- + u\, e^{-4\alpha - 8\beta_+} \right],
    \label{BII-ham-misner}
\end{equation}
It is possible to simplify Eq. (\ref{BII-ham-misner}) employing the canonical transformation \cite{Misner:1973prb}
\begin{equation}
    \begin{pmatrix}
        \alpha \\ \beta_+ \\ \beta_-
    \end{pmatrix}
    = \frac{1}{\sqrt{3}}
    \begin{pmatrix}
        2 & -1 & 0  \\
        -1 & 2 & 0  \\
        0 & 0 & 1
    \end{pmatrix}
    \begin{pmatrix}
        \tilde{\alpha} \\ \tilde{\beta}_+ \\ \tilde{\beta}_-
    \end{pmatrix}
    \,.
    \label{modified-coordinates}
\end{equation}
Since $\beta_-=\tilde{\beta}_-$, we will often omit the superscript $\sim$ to keep a lighter notation. With this particular choice of variables, the potential term appearing in Eq. (\ref{BII-ham-misner}) depends only on the new variable $\tilde{\beta}_+$, as the super-Hamiltonian constraint now reads as
\begin{equation}
        \frac{\chi e^{-3\alpha}}{3(8\pi)^2} \left( -\tilde{p}^2_\alpha + \tilde{p}^2_+ + p^2_- + u\, e^{-4\sqrt{3}\,\tilde{\beta}_+} \right)=0\,.
    \label{BII-ham-mod}
\end{equation}
Here $\tilde{p}_\alpha,\, \tilde{p}_+$ are the momenta conjugated to the new variables. From Eq. (\ref{BII-ham-mod}), we can notice that both $\tilde{p}_\alpha,\, p_-$ are constants of motion, while the evolution of $\tilde{\beta}_+$ is characterized by an elastic collision with the potential. The relation that links the anisotropy velocities of initial and final Kasner regimes is called \emph{BKL map}. For this set of coordinates, the map is rather trivial and reads as
\begin{equation}
    (\tilde{\beta}'_{+})_f = -(\tilde{\beta}'_{+})_i\,, \ (\beta'_{-})_f = (\beta'_{-})_i\,, \ (\tilde{p}_{\alpha})_f = (\tilde{p}_{\alpha})_i.
    \label{BKL-map-mod}
\end{equation}
Here, $i$ and $f$ label the Universe parameters before and after the collision. It is possible to check that the usual form of the BKL map is recovered using the transformation in Eq. (\ref{modified-coordinates}). Morevorer the relation $|(\bm{\beta}')_f|^2 = 1$ holds as long as $|(\bm{\beta}')_i|^2 = 1$. Therefore, the conditions needed for a new collision are always ensured.

\section{Bianchi II semiclassical evolution}
\begin{figure}[t]
    \centering
    \includegraphics[width=0.5\linewidth]{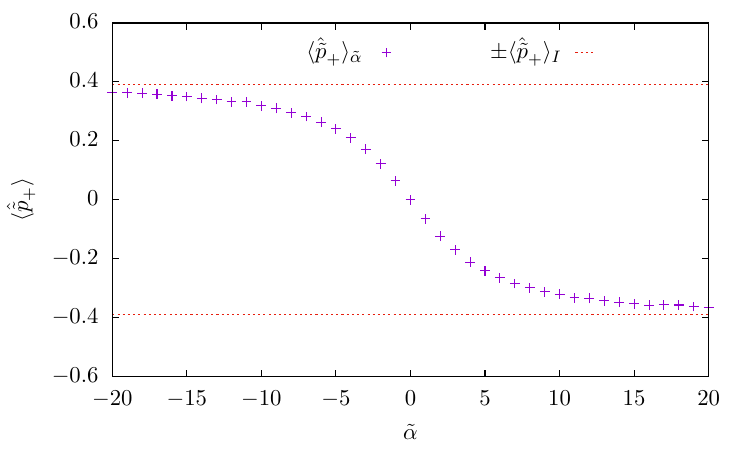}
    \caption{$\langle \hat{\tilde{p}}_+ \rangle_{\tilde{\alpha}}$ computed for a wave packet with $\mu_+ =-1,\, \sigma_+ =0.5,\,\mu_- =0,\, \sigma_- =1, u=1$ (purple points). The two red dashed lines represent the mean values computed over Bianchi I wave packets with $\langle \hat{\tilde{p}}_+\rangle_I \approx \pm 0.39$.}
    \label{Fig_1}
\end{figure}

We now move to the quantum mechanical framework and study the semiclassical evolution of the Bianchi II model. We rely on the Canonical approach to the quantization of gravity, and to be more precise, on the analogy between the Wheeler-DeWitt equation and the Klein-Gordon equation \cite{Wald_1993,Higuchi_1995}. This particular approach allows us to bypass the so-called \emph{problem of time} by selecting an internal degree of freedom as the time variable of the system \cite{rovelli_internal_time}. Generalized coordinates and conjugated momenta are promoted to operators acting on the so-called \emph{Universe wave functions} $\Psi(\alpha, \beta_\pm)$. In units $\hbar=c=G=1$, the super-Hamiltonian constraint (\ref{BII-ham-misner}) gives birth to the Wheeler-DeWitt equation
\begin{equation}
    \left[ \partial^2_\alpha - \partial^2_+ - \partial^2_- + u\, e^{-4\alpha-8\beta_+} \right] \Psi(\alpha, \beta_\pm)=0\,,
    \label{BII-wdw}
\end{equation}
where we denote the partial derivation with respect to $\beta_\pm$ as $\partial_\pm$. This equation is analogous to a Klein-Gordon equation in the presence of a potential, where the isotropic scale factor $\alpha$ has the role of a time variable. As in the classical case, we can further simplify the equation above by considering the coordinate transformation in Eq. (\ref{modified-coordinates}), which is a Lorentz transformation in the Mini-superspace and can be implemented without issues under the Lorentz invariance of the KG equation. Therefore, Eq. (\ref{BII-wdw}) now takes the form 
\begin{equation}
    \left[ \tilde{\partial}^2_\alpha - \tilde{\partial}^2_+ - \partial^2_- + u\, e^{-4\sqrt{3}\tilde{\beta}_+} \right] \Psi(\tilde{\alpha}, \tilde{\beta}_\pm)=0\,.
    \label{BII-wdw-mod}
\end{equation}
Similarly to the case of the closed FLRW model studied in \cite{kiefer-curv-sol,de_Cesare_2016,LoFranco:2024nss}, the potential involved in Eq. (\ref{BII-wdw-mod}) suggests imposing the boundary condition $\Psi(\tilde{\alpha}, \tilde{\beta}_\pm) \to 0$ for $\tilde{\beta}_+ \to -\infty$. Hence, the frequency-separated basis solutions of Eq. (\ref{BII-wdw-mod}) are 
\begin{equation}
    f^\pm_{\bf \tilde{k}} = \mathcal{N}_{\bf \tilde{k}}\, e^{\pm i\omega_{\bf \tilde{k}} \tilde{\alpha} + i k_- \beta_-} K_{\frac{i \tilde{k}_+}{2\sqrt{3}}} \left( \sqrt{\frac{u}{3}} \frac{e^{-2\sqrt{3}\,\tilde{\beta}_+}}{2} \right)
    \label{BII-wdw-mod-sol}
\end{equation}
where ${\bf \tilde{k}} = (\tilde{k}_+\,, k_-)$, $\omega_{\bf \tilde{k}} = \sqrt{\tilde{k}_+^2 + k_-^2}$, $K_\nu (z)$ are the modified Bessel functions of the second kind and $\mathcal{N}_{\bf \tilde{k}} = [\tilde{k}_+ \sinh{[\pi \tilde{k}_+/(2\sqrt{3})]}/(4\sqrt{3}\pi^3 \omega_{\bf \tilde{k}})]^{1/2}$ is the normalization factor. As the functions $K_\nu(z)$ are symmetric with respect to $\nu$, $\tilde{k}_+$ is constrained to only real positive values, while $k_- \in \mathbb{R}$. In the region where the potential $U_{II}$ is negligible, the solutions above take the form
\begin{equation}
    f^\pm_{\bf \tilde{k}} \approx \frac{e^{\pm i\omega_{\bf \tilde{k}} \tilde{\alpha} + i k_- \beta_-}}{\pi \sqrt{2 \omega_{\tilde{ \bf k}}}} \sin{\left[ \tilde{k}_+ \left( \tilde{\beta}_+ - C \right) - \gamma_{\tilde{k}_+} \right]}\,,
    \label{BII-wdw-mod-sol-lim}
\end{equation}
where $C\equiv \log[\sqrt{u}/(4 \sqrt{3})]/(2\sqrt{3})$ and $\gamma_\nu$ is a real and continuous parameter \cite{NIST:DLMF}. Therefore, the asymptotically free solutions are none other than a superposition of Bianchi I plane waves with opposite $\tilde{k}_+$. Given two functions $f,g$, we introduce the Klein-Gordon-like inner product for Eq. (\ref{BII-wdw-mod}) as
\begin{equation}
    \left( g\,,\, h \right) = -i\int d^2\tilde{\beta}\ g^*(\tilde{\alpha}, \tilde{\beta}_\pm) \overleftrightarrow{\tilde{\partial}_\alpha}\, h(\tilde{\alpha}, \tilde{\beta}_\pm)\,.
    \label{KG-inner-prod}
\end{equation}
Hence, the solutions in Eq. (\ref{BII-wdw-mod-sol}) satisfy the orthonormality relations
\begin{equation}
    \left( f^\pm_{\bf \tilde{k}'}\,,\, f^\pm_{\bf \tilde{k}} \right) = \pm \delta^{(2)}({\bf \tilde{k} - \tilde{k'}})\,, \quad \left( f^\pm_{\bf \tilde{k}'}\,,\, f^\mp_{\bf \tilde{k}} \right) = 0\,,
    \label{BII_orthonorm_rel}
\end{equation}
thanks to MacDonald functions' orthogonality \cite{PASSIAN2009380_macdonald}. Regarding the frequency separation of these solutions, the difference between positive and negative frequency solutions lies in the different direction of the internal time arrow (here represented by $\tilde{\alpha}$) that is followed during the semiclassical evolution \cite{LoFranco:2024nss}. However, the Bianchi II collision does not need to be represented as a quantum transition between opposite frequency solutions, as suggested by the fact that the potential in Eq. (\ref{BII-wdw-mod}) is independent of $\tilde{\alpha}$. Therefore, no frequency-mixing is allowed, and the quantum collision occurs inside the subsets of only positive/negative frequency solutions. From now on, we will discuss only positive frequency solutions, representing the current expansion Branch of the Universe. The first step towards deriving the semiclassical evolution of this model is to construct localized wave packets in the form 
\begin{equation}
    \Psi^+(\tilde{\alpha},\tilde{\beta}_\pm) = \int d^2 \tilde{k}\, A(\tilde{k}_\pm ; \mu_\pm, \sigma_\pm) f^+_{\bf \tilde{k}} (\tilde{\alpha},\tilde{\beta}_\pm)\,,
    \label{BII_wp}
\end{equation}
where the coefficients $A(\tilde{k}_\pm ; \mu_\pm, \sigma_\pm)$ are defined as
\begin{equation}
    A(\tilde{k}_\pm ; \mu_\pm, \sigma_\pm) = \frac{\exp{\left[ -\frac{(\log{\tilde{k}_+} - \mu_+)^2}{2\sigma_+^2}-\frac{(k_- - \mu_-)^2}{2\sigma_-^2}\right]}}{\sqrt{\tilde{k}_+ \sigma_+ \sigma_- \sqrt{\pi}}}\,.
    \label{gaussian-coefficients}
\end{equation}
The reasoning behind this particular form of the coefficients is that they satisfy the properties needed to construct well-normalized scattering amplitudes \cite{ISHIKAWA2024169571}. Therefore, we can introduce the probability density
\begin{equation}
    \mathcal{P}(\tilde{\alpha},\tilde{\beta}_\pm) = -i\ \Psi^{+*}(\tilde{\alpha}, \tilde{\beta}_\pm) \overleftrightarrow{\tilde{\partial}_\alpha}\, \Psi^+(\tilde{\alpha}, \tilde{\beta}_\pm)\,.
    \label{BII_pdens}
\end{equation} 
As long as the wave packets considered are only of positive frequency and properly normalized, the quantity above is indeed positive-definite. Evidence of the classical BKL map representing the semiclassical evolution of the Universe wave function can be provided from the behavior of the mean values of the momentum operators. Since the solutions in Eq. (\ref{BII-wdw-mod-sol}) are eigenstates of both $\hat{\tilde{p}}_\alpha\,,\, \hat{p}_-$, the mean values of such operators are constants in time and determined by $\mu_\pm,\sigma_\pm$. Regarding the operator $\hat{\tilde{p}}_+$, its mean value will depend on the internal time, and it can be computed using the probability density in Eq. (\ref{BII_pdens}), i.e. 
\begin{equation}
    \langle \hat{\tilde{p}}_+ \rangle_{\tilde{\alpha}} =  -i\int d^2\tilde{\beta}\ \Psi^{+*}(\tilde{\alpha}, \tilde{\beta}_\pm) \overleftrightarrow{\tilde{\partial}_\alpha}\, \left[ \hat{\tilde{p}}_+ \Psi^+(\tilde{\alpha}, \tilde{\beta}_\pm) \right].
    \label{BII_pp_mean}
\end{equation}
In Fig. \ref{Fig_1} we show the the evolution of $\langle \hat{\tilde{p}}_+ \rangle_{\tilde{\alpha}}$ with respect to the internal time $\tilde{\alpha}$. Denoting with $\langle \hat{\tilde{p}}_+ \rangle_{I}$ the mean value of $\hat{\tilde{p}}_+$ computed over a Bianchi I wave packet with same coefficients $A(\tilde{k}_\pm;\mu_\pm,\sigma_\pm)$, we see that $\langle \hat{\tilde{p}}_+ \rangle_{\tilde{\alpha}}$ starts at very early times with a value $\langle \hat{\tilde{p}}_+ \rangle_{I}$, and after the collision it will approach asymptotically the value $-\langle \hat{\tilde{p}}_+ \rangle_{I}$. This is indeed the reflection law characterizing the classical BKL map.
\section{The Bianchi II collision as a quantum scattering}
\begin{figure}[t]
    \begin{minipage}{0.49\columnwidth}
        \centering
        \includegraphics[width=1\columnwidth]{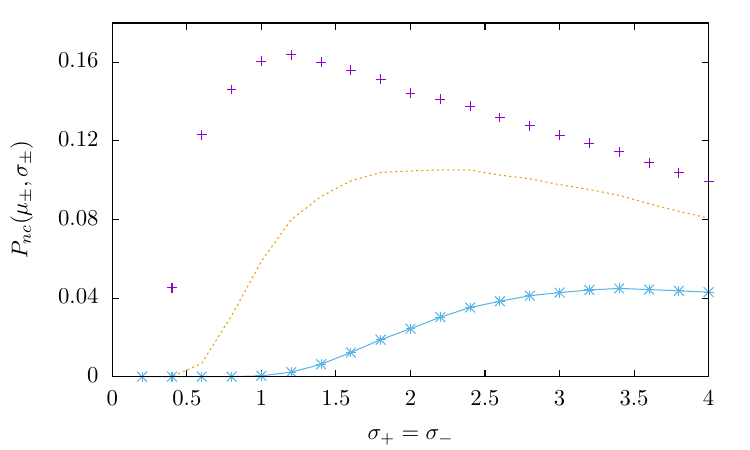}
        \caption{$P_{nc}(\mu_\pm,\sigma_\pm)$ as a function of the initial and final state $\sigma_+=\sigma_-$. Purple points: $\mu_+=-0.5,\mu_-=0.5$; yellow dashed line: $\mu_+=0,\mu_-=1$; blue points and line: $\mu_+=\mu_-=2$.}
        \label{Fig_2}
    \end{minipage} \hfill
    \begin{minipage}{0.49\columnwidth}
        \centering
        \includegraphics[width=1\columnwidth]{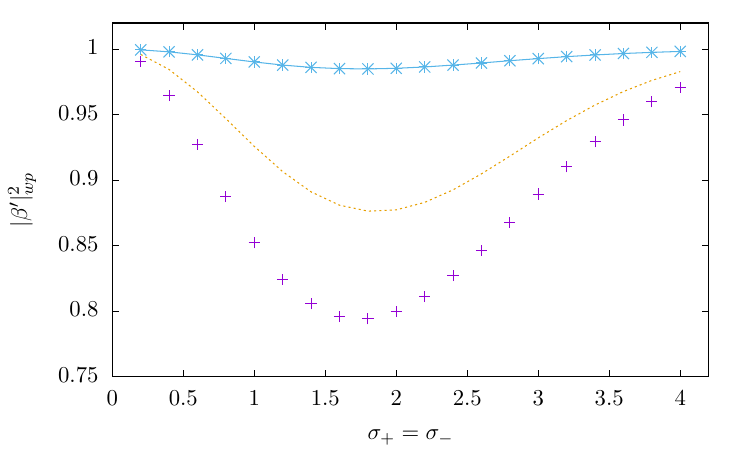}
        \caption{Incoming wave packet $|\bm \beta'|^2_{wp}$ as a function of $\sigma_+=\sigma_-$. Purple points: $\mu_+=-0.5,\mu_-=0.5$; yellow dashed line: $\mu_+=0,\mu_-=1$; blue points and line: $\mu_+=\mu_-=2$.} 
    \label{Fig_3}
    \end{minipage}
\end{figure}

We now study the collision mechanism of the Bianchi II model as a relativistic quantum scattering. %
We rely on the attempt to extend the WDW-KG analogy performed in \cite{Giovannetti:2022qje,giovannetti_maione,LoFranco:2024nss}, implementing the propagator scattering theory presented in \cite{Bjorken:Drell}. As we want to investigate the initial stages of the Universe's thermal history, we consider the scattering between positive frequency solutions propagating forward in time. Let us remark on the fact that the identification of the initial singularity at $\tilde{\alpha} \to +\infty$ is justified by the fact that after the Lorentz-like transformation in Eq. (\ref{modified-coordinates}), $\tilde{\alpha}$ plays the role of time variable as $\alpha$ do. The transition amplitude reads as \cite{Bjorken:Drell}
\begin{equation}
    \mathcal{S} = \lim_{\tilde{\alpha} \to -\infty} -i\int d^2\tilde{\beta}\ \psi^{+*}(\tilde{\alpha}, \tilde{\beta}_\pm) \overleftrightarrow{\tilde{\partial}_\alpha}\, \Psi^+(\tilde{\alpha}, \tilde{\beta}_\pm)\,.
    \label{S-amplitude-general}
\end{equation}
Here $\psi^+(\tilde{\alpha}, \tilde{\beta}_\pm)$ is a localized wave packet constructed as a superposition of positive frequency Bianchi I solutions, while $\Psi^+(\tilde{\alpha}, \tilde{\beta}_\pm)$ is a solution of the complete equation. In general, $\mathcal{S}$ will depend on the choice of parameters that shape the coefficients of in/out wavepackets (denoted respectively with $\mu_\pm,\sigma_\pm$ and $\mu'_\pm,\sigma'_\pm$). Since this outcoming state will be characterized by some value of $\langle \hat{\tilde{p}}_+ \rangle_{I}$, we can compute the probability of the Universe following the classical BKL map at the level of the momentum operators' mean values. Given our choice of wave packets coefficients and requiring that $\sigma'_\pm = \sigma_\pm$, $|\mathcal{S}|^2$ is a well normalized probability density (i.e., $\int d^2 \mu' |\mathcal{S}(\mu'_\pm;\mu_\pm)|^2 = 1$)  \cite{ISHIKAWA2024169571}. Now, to compute explicitly $|\mathcal{S}|^2$, we have to specify the form of the in/out states involved in the scattering. For the incoming state, we make an approximation suggested by the use of localized wave packets in the scattering amplitude. At the semiclassical level, we still have an evolution law valid for the mean value $\langle \hat{\tilde{\beta}}_+ \rangle_{\tilde{\alpha}}$ and we expect the Universe dynamics to be represented by Bianchi I solutions at arbitrarily large times $\tilde{\alpha}$. In this sense, we can say that large times $\tilde{\alpha}$ correspond to large values of the anisotropy parameter $\tilde{\beta}_+$. Therefore, we replace the limit $\tilde{\beta} \to +\infty$ in Eq. (\ref{S-amplitude-general}) so that our incoming state will be a superposition of solutions in Eq. (\ref{BII-wdw-mod-sol-lim}). Regarding the outcoming state, we consider a superposition of plane waves being characterized by a negative $\langle \hat{\tilde{p}}_+ \rangle_{I}$. Such a wave packet is in the form
\begin{equation}
    \psi^+ = \int d^2 \tilde{k}\, A(\tilde{k}_\pm ; \mu_\pm, \sigma_\pm) e^{+i\omega_{\tilde{\bf k}} \tilde{\alpha} + i {\bf \tilde{k} \cdot \tilde{\beta}} -i\, \gamma_{\tilde{k}_+}}, 
    \label{BI-wp} 
\end{equation}
Here the coefficients $A(\tilde{k}_\pm ; \mu_\pm, \sigma_\pm)$ are still the same of Eq. (\ref{gaussian-coefficients}), while $C$ and $\gamma_{\tilde{k}_+}$ are defined as in Eq. (\ref{BII-wdw-mod-sol-lim}). We added a $\tilde{k}_+$-dependent complex phase to simplify the phase term in Eq. (\ref{BII-wdw-mod-sol-lim}) when computing the scattering amplitude, as it does not affect the semiclassical evolution nor modifies the orthonormality relations. Moreover, we restrict the domain of $\tilde{k}_+$ to the negative half line to indeed constrain the value of $\langle \hat{\tilde{p}}_+ \rangle_{I}$ to only negative values. By substituting these states in Eq. (\ref{S-amplitude-general}), we find
\begin{equation}
    |\mathcal{S}(\mu'_\pm; \mu_\pm)|^2 = \frac{e^{-\frac{(\mu_+'-\mu_+)^2}{2\sigma_+^2} -\frac{(\mu_-'-\mu_-)^2}{2\sigma_-^2}}}{2\pi \sigma_+ \sigma_-}\,,
    \label{S-ammplitude-wp}
\end{equation}
that is a 2-dimensional Gaussian distribution. Therefore, the peak of the probability density is centered around $\mu'_\pm = \mu_\pm$, corresponding to the classical BKL reflection law. Although this is the most probable configuration, we ask ourselves if there is a non-negligible probability of the Universe exiting the scattering process in a state that won't collide again with $U_{IX}$. In the classical framework, the potential wall moves with an effective velocity $|\bm{\beta}'_{wall}|=1/2$ \cite{Montani:PrimCos}. 
Dealing with the quantum mechanical description, the classical quantities are replaced by mean values computed over their respective operators. Recalling that $\beta'_\pm = - p_\pm/p_\alpha$, the condition under which the quantum Universe collides against the potential reads as
\begin{equation}
    |\bm \beta'|^2_{wp} = \frac{(2 \langle\tilde{p}_+\rangle + \langle\tilde{p}_\alpha\rangle)^2 + 3\langle p^2_- \rangle}{(2\langle \tilde{p}_\alpha \rangle + \langle \tilde{p}_+\rangle)^2}\geq \frac{1}{4}\,.
    \label{kasnerianity}
\end{equation}
With Eq. (\ref{kasnerianity}), we can identify the subset of the parameter space $\mu'_\pm$ characterizing the solutions violating the classical BKL map. By integrating $|\mathcal{S}(\mu'_\pm; \mu_\pm)|^2$ on this subset, we can compute the probability $P_{nc}(\mu_\pm,\sigma_\pm)$ of the Universe transitioning to any of the BKL-violating states and study its behavior for fixed initial labels $\mu_\pm, \sigma_\pm=\sigma'_\pm$. As Eq. (\ref{kasnerianity}) is implemented for the outcoming states in Eq. (\ref{BI-wp}), it is important to unserdand the dependence of $\langle \hat{\tilde{p}}_\alpha\rangle_I,\langle \hat{\tilde{p}}_\pm\rangle_I$ on wave packets' parameters $\mu_\pm,\sigma_\pm$. As the $k_-$-dependent part of the coefficients $A(\tilde{k}_\pm ; \mu_\pm, \sigma_\pm)$ is a Gaussian distribution, the well-known relations $\langle\hat{p}_-\rangle_I=\mu_-,\,(\Delta \hat{p}_-)_I \propto \sigma_-$ hold. These dependencies are less trivial when considering the operator $\hat{\tilde{p}}_+$, since the $\tilde{k}_+$-dependent term of $A(\tilde{k}_\pm ; \mu_\pm, \sigma_\pm)$ is a log-normal distribution. Both $\mu_+,\sigma_+$ contributes to $\langle\hat{p}_+\rangle_I,\, (\Delta \hat{p}_+)_I$, while $\sigma_+$ alone fixes the value of the relative uncertainty $(\Delta \hat{p}_+/\langle\hat{p}_+\rangle)_I$. In particular, $\langle\hat{p}_+\rangle_I=-\exp[\mu_++\sigma_+^2/4]$. $\langle \hat{\tilde{p}}_\alpha\rangle_I$, instead, has been computed using numerical integrations. In Fig. (\ref{Fig_2}), we show the results of the computations for $P_{nc}(\mu_\pm,\sigma_\pm)$. For small values of $\sigma_\pm$, the probability of violating the classical BKL map is particularly low, and it increases for larger values of $\sigma_\pm$. This is a reasonable result as we expect low-variance wave packets to satisfy the condition $|\bm \beta'|_{wp}^2$. As the values of $\sigma_\pm$ increase more, the $P_{nc}$ reaches a maximum and starts decreasing again. Two arguments can clarify such behavior. The first is that large values of $\sigma_\pm$ can correspond to small values of the variance of the position operators $\tilde{\beta}_\pm$. Therefore, the classical BKL behavior is expected to be recovered at the level of mean values in the coordinate space. The second argument is that large values of $\sigma_+$ also correspond to large values of $\langle \hat{\tilde{p}}_+ \rangle_I$. Hence, we recover Minser's result of high occupation numbers states (here represented by high-momenta states) following the classical dynamics \cite{misner_qc1}. This argument is also backed by the behavior of $P_{nc}$ when varying $\mu_\pm$. We can see that for large values of $\mu_\pm$, corresponding to large values of $\langle \hat{\tilde{p}}_\pm \rangle_I$, the probability of BKL-violations is reduced drastically. However, in the region where neither the momenta are well localized nor the total momentum is high enough, $P_{nc}$ assumes non-negligible values. To better clarify these results, in Fig. \ref{Fig_3}, we show the values of $|\bm \beta'|^2_{wp}$ computed over Bianchi I wave packets representing the initial state. As we can see, the behavior with respect to $\mu_\pm,\sigma_\pm$ is the same as Fig. \ref{Fig_2}, i.e., $|\bm \beta'|^2_{wp} \to 1$ for very small/large values of $\sigma_\pm$ and large values of $\mu_\pm$. Therefore, the more classical the initial state is, the less the probability of violating the BKL map.       
\section{Concluding remarks}
We analyzed the quantum Mixmaster model using the formalism typical of the relativistic quantum scattering \cite{Bjorken:Drell}, basing our formulation on the pseudo-Riemannian character of the Mini-supermetric. We could construct a Klein-Gordon-like representation of the Wheeler-DeWitt equation for the Bianchi IX cosmology, which allows the construction of suitable Hilbert spaces \cite{Wald_1993} and the determination of transition amplitudes among states. In particular, we treated the bouncing of the pin-point particle against a single wall (i.e., the exact Bianchi II dynamics) as a quantum scattering onto that potential. Thus, we could calculate the transition amplitude from an initial toward a final Kasner regime, including states for which a subsequent bounce is forbidden, i.e., the quantum transition violates the BKL map \cite{BKL_1982}, though it remains valid for the mean values. The most important result here is the validation of Misner's original idea that quasi-classical states can survive up to the initial singularity. In fact, very localized initial states in the momentum space (whose uncertainty remains constant in time) have very negligible transition amplitudes with states violating the BKL map. This scheme can be iterated for an arbitrarily large number of bounces toward the singularity, and we thus deal with quantum states that are indefinitely following the semi-classical BKL map. This situation also holds for very large values of the momentum uncertainties, corresponding to very localized states in the anisotropies. However, in such a representation, the possibility of iterating the picture is unclear since the wave packets naturally spread in the anisotropy space. However, the high occupation number, discussed by Misner \cite{misner_qc1}, clearly corresponds to dealing with high mean values of the anisotropy momenta, and also in this case, the probability of violating the BKL map is negligible. It is just this result that sheds light on the concept of semiclassical states approaching the initial singularity. Furthermore, when the momentum uncertainty of the initial state is large enough, the transition toward no longer bouncing configurations, as described by the average Kasner regime, takes significantly non-zero values. This result could align with the reduction of chaoticity presented in \cite{Bojowald:2023fas,Bojowald:2023sjw,Brizuela:2022uun,Uria:2023uik}. This is an important issue that clarifies how the quantum Mixmaster can be a very different system from a quasi-classical chaotic dynamics, as soon the real quantum nature of the pin-point particle is taken into account. 
\bibliography{references}

\begin{thebibliography}{38}%
\makeatletter
\providecommand \@ifxundefined [1]{%
 \@ifx{#1\undefined}
}%
\providecommand \@ifnum [1]{%
 \ifnum #1\expandafter \@firstoftwo
 \else \expandafter \@secondoftwo
 \fi
}%
\providecommand \@ifx [1]{%
 \ifx #1\expandafter \@firstoftwo
 \else \expandafter \@secondoftwo
 \fi
}%
\providecommand \natexlab [1]{#1}%
\providecommand \enquote  [1]{``#1''}%
\providecommand \bibnamefont  [1]{#1}%
\providecommand \bibfnamefont [1]{#1}%
\providecommand \citenamefont [1]{#1}%
\providecommand \href@noop [0]{\@secondoftwo}%
\providecommand \href [0]{\begingroup \@sanitize@url \@href}%
\providecommand \@href[1]{\@@startlink{#1}\@@href}%
\providecommand \@@href[1]{\endgroup#1\@@endlink}%
\providecommand \@sanitize@url [0]{\catcode `\\12\catcode `\$12\catcode `\&12\catcode `\#12\catcode `\^12\catcode `\_12\catcode `\%12\relax}%
\providecommand \@@startlink[1]{}%
\providecommand \@@endlink[0]{}%
\providecommand \url  [0]{\begingroup\@sanitize@url \@url }%
\providecommand \@url [1]{\endgroup\@href {#1}{\urlprefix }}%
\providecommand \urlprefix  [0]{URL }%
\providecommand \Eprint [0]{\href }%
\providecommand \doibase [0]{https://doi.org/}%
\providecommand \selectlanguage [0]{\@gobble}%
\providecommand \bibinfo  [0]{\@secondoftwo}%
\providecommand \bibfield  [0]{\@secondoftwo}%
\providecommand \translation [1]{[#1]}%
\providecommand \BibitemOpen [0]{}%
\providecommand \bibitemStop [0]{}%
\providecommand \bibitemNoStop [0]{.\EOS\space}%
\providecommand \EOS [0]{\spacefactor3000\relax}%
\providecommand \BibitemShut  [1]{\csname bibitem#1\endcsname}%
\let\auto@bib@innerbib\@empty
\bibitem [{\citenamefont {Belinskiĭ}\ \emph {et~al.}(1971)\citenamefont {Belinskiĭ}, \citenamefont {Lifshitz},\ and\ \citenamefont {Khalatnikov}}]{BKL_1971}%
  \BibitemOpen
  \bibfield  {author} {\bibinfo {author} {\bibfnamefont {V.~A.}\ \bibnamefont {Belinskiĭ}}, \bibinfo {author} {\bibfnamefont {E.~M.}\ \bibnamefont {Lifshitz}},\ and\ \bibinfo {author} {\bibfnamefont {I.~M.}\ \bibnamefont {Khalatnikov}},\ }\bibfield  {title} {\bibinfo {title} {Oscillatory approach to the singular point in relativistic cosmology},\ }\href {https://doi.org/10.1070/PU1971v013n06ABEH004279} {\bibfield  {journal} {\bibinfo  {journal} {Soviet Physics Uspekhi}\ }\textbf {\bibinfo {volume} {13}},\ \bibinfo {pages} {745} (\bibinfo {year} {1971})}\BibitemShut {NoStop}%
\bibitem [{\citenamefont {Montani}\ \emph {et~al.}(2009)\citenamefont {Montani}, \citenamefont {Battisti}, \citenamefont {Benini},\ and\ \citenamefont {Imponente}}]{Montani:PrimCos}%
  \BibitemOpen
  \bibfield  {author} {\bibinfo {author} {\bibfnamefont {G.}~\bibnamefont {Montani}}, \bibinfo {author} {\bibfnamefont {M.~V.}\ \bibnamefont {Battisti}}, \bibinfo {author} {\bibfnamefont {R.}~\bibnamefont {Benini}},\ and\ \bibinfo {author} {\bibfnamefont {G.}~\bibnamefont {Imponente}},\ }\href@noop {} {\emph {\bibinfo {title} {{Primordial cosmology}}}}\ (\bibinfo  {publisher} {World Scientific},\ \bibinfo {address} {Singapore},\ \bibinfo {year} {2009})\BibitemShut {NoStop}%
\bibitem [{\citenamefont {Montani}\ \emph {et~al.}(2008)\citenamefont {Montani}, \citenamefont {Battisti}, \citenamefont {Benini},\ and\ \citenamefont {Imponente}}]{montani_mix_review}%
  \BibitemOpen
  \bibfield  {author} {\bibinfo {author} {\bibfnamefont {G.}~\bibnamefont {Montani}}, \bibinfo {author} {\bibfnamefont {M.~V.}\ \bibnamefont {Battisti}}, \bibinfo {author} {\bibfnamefont {R.}~\bibnamefont {Benini}},\ and\ \bibinfo {author} {\bibfnamefont {G.}~\bibnamefont {Imponente}},\ }\bibfield  {title} {\bibinfo {title} {Classical and quantum features of the mixmaster singularity},\ }\href {https://doi.org/10.1142/S0217751X08040275} {\bibfield  {journal} {\bibinfo  {journal} {International Journal of Modern Physics A}\ }\textbf {\bibinfo {volume} {23}},\ \bibinfo {pages} {2353} (\bibinfo {year} {2008})},\ \Eprint {https://arxiv.org/abs/https://doi.org/10.1142/S0217751X08040275} {https://doi.org/10.1142/S0217751X08040275} \BibitemShut {NoStop}%
\bibitem [{\citenamefont {Belinsky}\ \emph {et~al.}(1982)\citenamefont {Belinsky}, \citenamefont {Khalatnikov},\ and\ \citenamefont {Lifshitz}}]{BKL_1982}%
  \BibitemOpen
  \bibfield  {author} {\bibinfo {author} {\bibfnamefont {V.~A.}\ \bibnamefont {Belinsky}}, \bibinfo {author} {\bibfnamefont {I.~M.}\ \bibnamefont {Khalatnikov}},\ and\ \bibinfo {author} {\bibfnamefont {E.~M.}\ \bibnamefont {Lifshitz}},\ }\bibfield  {title} {\bibinfo {title} {{A General Solution of the Einstein Equations with a Time Singularity}},\ }\href {https://doi.org/10.1080/00018738200101428} {\bibfield  {journal} {\bibinfo  {journal} {Adv. Phys.}\ }\textbf {\bibinfo {volume} {31}},\ \bibinfo {pages} {639} (\bibinfo {year} {1982})}\BibitemShut {NoStop}%
\bibitem [{\citenamefont {Kirillov}(1993)}]{kirillov_93}%
  \BibitemOpen
  \bibfield  {author} {\bibinfo {author} {\bibfnamefont {A.}~\bibnamefont {Kirillov}},\ }\bibfield  {title} {\bibinfo {title} {On the nature of the spatial distribution of metric inhomogeneities in the general solution of the einstein equations near a cosmological singularity},\ }\href@noop {} {\bibfield  {journal} {\bibinfo  {journal} {Sov. Phys. JETP}\ }\textbf {\bibinfo {volume} {76}} (\bibinfo {year} {1993})}\BibitemShut {NoStop}%
\bibitem [{\citenamefont {Montani}(1995)}]{Montani_1995}%
  \BibitemOpen
  \bibfield  {author} {\bibinfo {author} {\bibfnamefont {G.}~\bibnamefont {Montani}},\ }\bibfield  {title} {\bibinfo {title} {On the general behaviour of the universe near the cosmological singularity},\ }\href {https://doi.org/10.1088/0264-9381/12/10/010} {\bibfield  {journal} {\bibinfo  {journal} {Classical and Quantum Gravity}\ }\textbf {\bibinfo {volume} {12}},\ \bibinfo {pages} {2505} (\bibinfo {year} {1995})}\BibitemShut {NoStop}%
\bibitem [{\citenamefont {{Belinskij}}(1992)}]{BL_1992}%
  \BibitemOpen
  \bibfield  {author} {\bibinfo {author} {\bibfnamefont {V.~A.}\ \bibnamefont {{Belinskij}}},\ }\bibfield  {title} {\bibinfo {title} {{Turbulence of a gravitational field near a cosmological singularity.}},\ }\href@noop {} {\bibfield  {journal} {\bibinfo  {journal} {Soviet Journal of Experimental and Theoretical Physics Letters}\ }\textbf {\bibinfo {volume} {56}},\ \bibinfo {pages} {421} (\bibinfo {year} {1992})}\BibitemShut {NoStop}%
\bibitem [{\citenamefont {Barrow}(2020)}]{Barrow:2020rsp}%
  \BibitemOpen
  \bibfield  {author} {\bibinfo {author} {\bibfnamefont {J.~D.}\ \bibnamefont {Barrow}},\ }\bibfield  {title} {\bibinfo {title} {{Multifractality in the general cosmological solution of Einstein\textquoteright{}s equations}},\ }\href {https://doi.org/10.1103/PhysRevD.102.041501} {\bibfield  {journal} {\bibinfo  {journal} {Phys. Rev. D}\ }\textbf {\bibinfo {volume} {102}},\ \bibinfo {pages} {041501} (\bibinfo {year} {2020})},\ \Eprint {https://arxiv.org/abs/2006.08652} {arXiv:2006.08652 [gr-qc]} \BibitemShut {NoStop}%
\bibitem [{\citenamefont {Misner}(1969{\natexlab{a}})}]{Misner_mixmaster68}%
  \BibitemOpen
  \bibfield  {author} {\bibinfo {author} {\bibfnamefont {C.~W.}\ \bibnamefont {Misner}},\ }\bibfield  {title} {\bibinfo {title} {{Mixmaster universe}},\ }\href {https://doi.org/10.1103/PhysRevLett.22.1071} {\bibfield  {journal} {\bibinfo  {journal} {Phys. Rev. Lett.}\ }\textbf {\bibinfo {volume} {22}},\ \bibinfo {pages} {1071} (\bibinfo {year} {1969}{\natexlab{a}})}\BibitemShut {NoStop}%
\bibitem [{\citenamefont {Cornish}\ and\ \citenamefont {Levin}(1997)}]{cornish_levin_97}%
  \BibitemOpen
  \bibfield  {author} {\bibinfo {author} {\bibfnamefont {N.~J.}\ \bibnamefont {Cornish}}\ and\ \bibinfo {author} {\bibfnamefont {J.~J.}\ \bibnamefont {Levin}},\ }\bibfield  {title} {\bibinfo {title} {The mixmaster universe is chaotic},\ }\href {https://doi.org/10.1103/PhysRevLett.78.998} {\bibfield  {journal} {\bibinfo  {journal} {Phys. Rev. Lett.}\ }\textbf {\bibinfo {volume} {78}},\ \bibinfo {pages} {998} (\bibinfo {year} {1997})}\BibitemShut {NoStop}%
\bibitem [{\citenamefont {Imponente}\ and\ \citenamefont {Montani}(2001)}]{montani_imponente_01}%
  \BibitemOpen
  \bibfield  {author} {\bibinfo {author} {\bibfnamefont {G.}~\bibnamefont {Imponente}}\ and\ \bibinfo {author} {\bibfnamefont {G.}~\bibnamefont {Montani}},\ }\bibfield  {title} {\bibinfo {title} {Covariance of the mixmaster chaoticity},\ }\href {https://doi.org/10.1103/PhysRevD.63.103501} {\bibfield  {journal} {\bibinfo  {journal} {Phys. Rev. D}\ }\textbf {\bibinfo {volume} {63}},\ \bibinfo {pages} {103501} (\bibinfo {year} {2001})}\BibitemShut {NoStop}%
\bibitem [{\citenamefont {Benini}\ and\ \citenamefont {Montani}(2004)}]{montani_benini_04}%
  \BibitemOpen
  \bibfield  {author} {\bibinfo {author} {\bibfnamefont {R.}~\bibnamefont {Benini}}\ and\ \bibinfo {author} {\bibfnamefont {G.}~\bibnamefont {Montani}},\ }\bibfield  {title} {\bibinfo {title} {Frame independence of the inhomogeneous mixmaster chaos via misner-chitr\'e-like variables},\ }\href {https://doi.org/10.1103/PhysRevD.70.103527} {\bibfield  {journal} {\bibinfo  {journal} {Phys. Rev. D}\ }\textbf {\bibinfo {volume} {70}},\ \bibinfo {pages} {103527} (\bibinfo {year} {2004})}\BibitemShut {NoStop}%
\bibitem [{\citenamefont {DeWitt}(1967)}]{DW_QC_1}%
  \BibitemOpen
  \bibfield  {author} {\bibinfo {author} {\bibfnamefont {B.~S.}\ \bibnamefont {DeWitt}},\ }\bibfield  {title} {\bibinfo {title} {Quantum theory of gravity. {I}. the canonical theory},\ }\href {https://doi.org/10.1103/PhysRev.160.1113} {\bibfield  {journal} {\bibinfo  {journal} {Phys. Rev.}\ }\textbf {\bibinfo {volume} {160}},\ \bibinfo {pages} {1113} (\bibinfo {year} {1967})}\BibitemShut {NoStop}%
\bibitem [{\citenamefont {DeWitt}(1968{\natexlab{a}})}]{DW_QC_2}%
  \BibitemOpen
  \bibfield  {author} {\bibinfo {author} {\bibfnamefont {B.~S.}\ \bibnamefont {DeWitt}},\ }\bibfield  {title} {\bibinfo {title} {Quantum theory of gravity. {II}. the manifestly covariant theory},\ }\href {https://doi.org/10.1103/PhysRev.171.1834.3} {\bibfield  {journal} {\bibinfo  {journal} {Phys. Rev.}\ }\textbf {\bibinfo {volume} {171}},\ \bibinfo {pages} {1834} (\bibinfo {year} {1968}{\natexlab{a}})}\BibitemShut {NoStop}%
\bibitem [{\citenamefont {DeWitt}(1968{\natexlab{b}})}]{DW_QC_3}%
  \BibitemOpen
  \bibfield  {author} {\bibinfo {author} {\bibfnamefont {B.~S.}\ \bibnamefont {DeWitt}},\ }\bibfield  {title} {\bibinfo {title} {Quantum theory of gravity. {III}. applications of the covariant theory},\ }\href {https://doi.org/10.1103/PhysRev.171.1834.4} {\bibfield  {journal} {\bibinfo  {journal} {Phys. Rev.}\ }\textbf {\bibinfo {volume} {171}},\ \bibinfo {pages} {1834} (\bibinfo {year} {1968}{\natexlab{b}})}\BibitemShut {NoStop}%
\bibitem [{\citenamefont {Misner}(1969{\natexlab{b}})}]{misner_qc1}%
  \BibitemOpen
  \bibfield  {author} {\bibinfo {author} {\bibfnamefont {C.~W.}\ \bibnamefont {Misner}},\ }\bibfield  {title} {\bibinfo {title} {Quantum cosmology. {I}},\ }\href {https://doi.org/10.1103/PhysRev.186.1319} {\bibfield  {journal} {\bibinfo  {journal} {Phys. Rev.}\ }\textbf {\bibinfo {volume} {186}},\ \bibinfo {pages} {1319} (\bibinfo {year} {1969}{\natexlab{b}})}\BibitemShut {NoStop}%
\bibitem [{\citenamefont {Benini}\ and\ \citenamefont {Montani}(2006)}]{montani_benini2006inhomogeneous}%
  \BibitemOpen
  \bibfield  {author} {\bibinfo {author} {\bibfnamefont {R.}~\bibnamefont {Benini}}\ and\ \bibinfo {author} {\bibfnamefont {G.}~\bibnamefont {Montani}},\ }\bibfield  {title} {\bibinfo {title} {Inhomogeneous quantum mixmaster: From classical towards quantum mechanics},\ }\href@noop {} {\bibfield  {journal} {\bibinfo  {journal} {Classical and Quantum Gravity}\ }\textbf {\bibinfo {volume} {24}},\ \bibinfo {pages} {387} (\bibinfo {year} {2006})}\BibitemShut {NoStop}%
\bibitem [{\citenamefont {Antonini}\ and\ \citenamefont {Montani}(2019)}]{montani_antonini2019singularity}%
  \BibitemOpen
  \bibfield  {author} {\bibinfo {author} {\bibfnamefont {S.}~\bibnamefont {Antonini}}\ and\ \bibinfo {author} {\bibfnamefont {G.}~\bibnamefont {Montani}},\ }\bibfield  {title} {\bibinfo {title} {Singularity-free and non-chaotic inhomogeneous mixmaster in polymer representation for the volume of the universe},\ }\href@noop {} {\bibfield  {journal} {\bibinfo  {journal} {Physics Letters B}\ }\textbf {\bibinfo {volume} {790}},\ \bibinfo {pages} {475} (\bibinfo {year} {2019})}\BibitemShut {NoStop}%
\bibitem [{\citenamefont {Wald}(1993)}]{Wald_1993}%
  \BibitemOpen
  \bibfield  {author} {\bibinfo {author} {\bibfnamefont {R.~M.}\ \bibnamefont {Wald}},\ }\bibfield  {title} {\bibinfo {title} {Proposal for solving the `problem of time' in canonical quantum gravity},\ }\href {https://doi.org/10.1103/physrevd.48.r2377} {\bibfield  {journal} {\bibinfo  {journal} {Physical Review D}\ }\textbf {\bibinfo {volume} {48}},\ \bibinfo {pages} {R2377} (\bibinfo {year} {1993})}\BibitemShut {NoStop}%
\bibitem [{\citenamefont {Higuchi}\ and\ \citenamefont {Wald}(1995)}]{Higuchi_1995}%
  \BibitemOpen
  \bibfield  {author} {\bibinfo {author} {\bibfnamefont {A.}~\bibnamefont {Higuchi}}\ and\ \bibinfo {author} {\bibfnamefont {R.~M.}\ \bibnamefont {Wald}},\ }\bibfield  {title} {\bibinfo {title} {Applications of a new proposal for solving the `problem of time' to some simple quantum cosmological models},\ }\href {https://doi.org/10.1103/physrevd.51.544} {\bibfield  {journal} {\bibinfo  {journal} {Physical Review D}\ }\textbf {\bibinfo {volume} {51}},\ \bibinfo {pages} {544} (\bibinfo {year} {1995})}\BibitemShut {NoStop}%
\bibitem [{\citenamefont {Cianfrani}\ \emph {et~al.}(2014)\citenamefont {Cianfrani}, \citenamefont {Lecian}, \citenamefont {Lulli},\ and\ \citenamefont {Montani}}]{montani_cqg}%
  \BibitemOpen
  \bibfield  {author} {\bibinfo {author} {\bibfnamefont {F.}~\bibnamefont {Cianfrani}}, \bibinfo {author} {\bibfnamefont {O.~M.}\ \bibnamefont {Lecian}}, \bibinfo {author} {\bibfnamefont {M.}~\bibnamefont {Lulli}},\ and\ \bibinfo {author} {\bibfnamefont {G.}~\bibnamefont {Montani}},\ }\href {https://doi.org/10.1142/8957} {\emph {\bibinfo {title} {Canonical Quantum Gravity}}}\ (\bibinfo  {publisher} {WORLD SCIENTIFIC},\ \bibinfo {year} {2014})\ \Eprint {https://arxiv.org/abs/https://www.worldscientific.com/doi/pdf/10.1142/8957} {https://www.worldscientific.com/doi/pdf/10.1142/8957} \BibitemShut {NoStop}%
\bibitem [{\citenamefont {Giovannetti}\ and\ \citenamefont {Montani}(2022)}]{Giovannetti:2022qje}%
  \BibitemOpen
  \bibfield  {author} {\bibinfo {author} {\bibfnamefont {E.}~\bibnamefont {Giovannetti}}\ and\ \bibinfo {author} {\bibfnamefont {G.}~\bibnamefont {Montani}},\ }\bibfield  {title} {\bibinfo {title} {{Is Bianchi I a bouncing cosmology in the Wheeler-DeWitt picture?}},\ }\href {https://doi.org/10.1103/PhysRevD.106.044053} {\bibfield  {journal} {\bibinfo  {journal} {Phys. Rev. D}\ }\textbf {\bibinfo {volume} {106}},\ \bibinfo {pages} {044053} (\bibinfo {year} {2022})},\ \Eprint {https://arxiv.org/abs/2203.01062} {arXiv:2203.01062 [gr-qc]} \BibitemShut {NoStop}%
\bibitem [{\citenamefont {Giovannetti}\ \emph {et~al.}(2023)\citenamefont {Giovannetti}, \citenamefont {Maione},\ and\ \citenamefont {Montani}}]{giovannetti_maione}%
  \BibitemOpen
  \bibfield  {author} {\bibinfo {author} {\bibfnamefont {E.}~\bibnamefont {Giovannetti}}, \bibinfo {author} {\bibfnamefont {F.}~\bibnamefont {Maione}},\ and\ \bibinfo {author} {\bibfnamefont {G.}~\bibnamefont {Montani}},\ }\bibfield  {title} {\bibinfo {title} {Quantum big bounce of the isotropic universe using relational time},\ }\bibfield  {journal} {\bibinfo  {journal} {Universe}\ }\textbf {\bibinfo {volume} {9}},\ \href {https://doi.org/10.3390/universe9080373} {10.3390/universe9080373} (\bibinfo {year} {2023})\BibitemShut {NoStop}%
\bibitem [{\citenamefont {Lo~Franco}\ and\ \citenamefont {Montani}(2024)}]{LoFranco:2024nss}%
  \BibitemOpen
  \bibfield  {author} {\bibinfo {author} {\bibfnamefont {S.}~\bibnamefont {Lo~Franco}}\ and\ \bibinfo {author} {\bibfnamefont {G.}~\bibnamefont {Montani}},\ }\bibfield  {title} {\bibinfo {title} {{Quantum Big-Bounce as a phenomenology of RQM in the Mini-superspace}},\ }\href {https://doi.org/10.1016/j.physletb.2024.138983} {\bibfield  {journal} {\bibinfo  {journal} {Phys. Lett. B}\ }\textbf {\bibinfo {volume} {857}},\ \bibinfo {pages} {138983} (\bibinfo {year} {2024})},\ \Eprint {https://arxiv.org/abs/2404.02802} {arXiv:2404.02802 [gr-qc]} \BibitemShut {NoStop}%
\bibitem [{\citenamefont {Bjorken}\ and\ \citenamefont {Drell}(1964)}]{Bjorken:Drell}%
  \BibitemOpen
  \bibfield  {author} {\bibinfo {author} {\bibfnamefont {J.~D.}\ \bibnamefont {Bjorken}}\ and\ \bibinfo {author} {\bibfnamefont {S.~D.}\ \bibnamefont {Drell}},\ }\href {https://cds.cern.ch/record/100769} {\emph {\bibinfo {title} {{Relativistic quantum mechanics}}}},\ International series in pure and applied physics\ (\bibinfo  {publisher} {McGraw-Hill},\ \bibinfo {address} {New York, NY},\ \bibinfo {year} {1964})\BibitemShut {NoStop}%
\bibitem [{\citenamefont {Ashtekar}\ \emph {et~al.}(2010)\citenamefont {Ashtekar}, \citenamefont {Alimi},\ and\ \citenamefont {Fu{\"o}zfa}}]{Ashtekar_2010}%
  \BibitemOpen
  \bibfield  {author} {\bibinfo {author} {\bibfnamefont {A.}~\bibnamefont {Ashtekar}}, \bibinfo {author} {\bibfnamefont {J.-M.}\ \bibnamefont {Alimi}},\ and\ \bibinfo {author} {\bibfnamefont {A.}~\bibnamefont {Fu{\"o}zfa}},\ }\bibfield  {title} {\bibinfo {title} {The big bang and the quantum},\ }in\ \href {https://doi.org/10.1063/1.3462605} {\emph {\bibinfo {booktitle} {AIP Conference Proceedings}}}\ (\bibinfo  {publisher} {AIP},\ \bibinfo {year} {2010})\ p.\ \bibinfo {pages} {109–121}\BibitemShut {NoStop}%
\bibitem [{\citenamefont {Barca}\ \emph {et~al.}(2021)\citenamefont {Barca}, \citenamefont {Giovannetti},\ and\ \citenamefont {Montani}}]{Barca_2021}%
  \BibitemOpen
  \bibfield  {author} {\bibinfo {author} {\bibfnamefont {G.}~\bibnamefont {Barca}}, \bibinfo {author} {\bibfnamefont {E.}~\bibnamefont {Giovannetti}},\ and\ \bibinfo {author} {\bibfnamefont {G.}~\bibnamefont {Montani}},\ }\bibfield  {title} {\bibinfo {title} {An overview on the nature of the bounce in {LQC} and {PQM}},\ }\href {https://doi.org/10.3390/universe7090327} {\bibfield  {journal} {\bibinfo  {journal} {Universe}\ }\textbf {\bibinfo {volume} {7}},\ \bibinfo {pages} {327} (\bibinfo {year} {2021})}\BibitemShut {NoStop}%
\bibitem [{\citenamefont {Misner}\ \emph {et~al.}(1973)\citenamefont {Misner}, \citenamefont {Thorne},\ and\ \citenamefont {Wheeler}}]{Misner:1973prb}%
  \BibitemOpen
  \bibfield  {author} {\bibinfo {author} {\bibfnamefont {C.~W.}\ \bibnamefont {Misner}}, \bibinfo {author} {\bibfnamefont {K.~S.}\ \bibnamefont {Thorne}},\ and\ \bibinfo {author} {\bibfnamefont {J.~A.}\ \bibnamefont {Wheeler}},\ }\href@noop {} {\emph {\bibinfo {title} {{Gravitation}}}}\ (\bibinfo  {publisher} {W. H. Freeman},\ \bibinfo {address} {San Francisco},\ \bibinfo {year} {1973})\BibitemShut {NoStop}%
\bibitem [{\citenamefont {Rovelli}(1991)}]{rovelli_internal_time}%
  \BibitemOpen
  \bibfield  {author} {\bibinfo {author} {\bibfnamefont {C.}~\bibnamefont {Rovelli}},\ }\bibfield  {title} {\bibinfo {title} {Time in quantum gravity: An hypothesis},\ }\href {https://doi.org/10.1103/PhysRevD.43.442} {\bibfield  {journal} {\bibinfo  {journal} {Phys. Rev. D}\ }\textbf {\bibinfo {volume} {43}},\ \bibinfo {pages} {442} (\bibinfo {year} {1991})}\BibitemShut {NoStop}%
\bibitem [{\citenamefont {Kiefer}(1988)}]{kiefer-curv-sol}%
  \BibitemOpen
  \bibfield  {author} {\bibinfo {author} {\bibfnamefont {C.}~\bibnamefont {Kiefer}},\ }\bibfield  {title} {\bibinfo {title} {Wave packets in minisuperspace},\ }\href {https://doi.org/10.1103/PhysRevD.38.1761} {\bibfield  {journal} {\bibinfo  {journal} {Phys. Rev. D}\ }\textbf {\bibinfo {volume} {38}},\ \bibinfo {pages} {1761} (\bibinfo {year} {1988})}\BibitemShut {NoStop}%
\bibitem [{\citenamefont {de~Cesare}\ \emph {et~al.}(2016)\citenamefont {de~Cesare}, \citenamefont {Gargiulo},\ and\ \citenamefont {Sakellariadou}}]{de_Cesare_2016}%
  \BibitemOpen
  \bibfield  {author} {\bibinfo {author} {\bibfnamefont {M.}~\bibnamefont {de~Cesare}}, \bibinfo {author} {\bibfnamefont {M.~V.}\ \bibnamefont {Gargiulo}},\ and\ \bibinfo {author} {\bibfnamefont {M.}~\bibnamefont {Sakellariadou}},\ }\bibfield  {title} {\bibinfo {title} {Semiclassical solutions of generalized wheeler-dewitt cosmology},\ }\bibfield  {journal} {\bibinfo  {journal} {Physical Review D}\ }\textbf {\bibinfo {volume} {93}},\ \href {https://doi.org/10.1103/physrevd.93.024046} {10.1103/physrevd.93.024046} (\bibinfo {year} {2016})\BibitemShut {NoStop}%
\bibitem [{{\relax DLMF}()}]{NIST:DLMF}%
  \BibitemOpen
  {\relax DLMF},\ \href {https://dlmf.nist.gov/} {\bibinfo {title} {{\it NIST Digital Library of Mathematical Functions}}},\ \bibinfo {howpublished} {\url{https://dlmf.nist.gov/}, Release 1.2.3 of 2024-12-15},\ \bibinfo {note} {f.~W.~J. Olver, A.~B. {Olde Daalhuis}, D.~W. Lozier, B.~I. Schneider, R.~F. Boisvert, C.~W. Clark, B.~R. Miller, B.~V. Saunders, H.~S. Cohl, and M.~A. McClain, eds.}\BibitemShut {Stop}%
\bibitem [{\citenamefont {Passian}\ \emph {et~al.}(2009)\citenamefont {Passian}, \citenamefont {Simpson}, \citenamefont {Kouchekian},\ and\ \citenamefont {Yakubovich}}]{PASSIAN2009380_macdonald}%
  \BibitemOpen
  \bibfield  {author} {\bibinfo {author} {\bibfnamefont {A.}~\bibnamefont {Passian}}, \bibinfo {author} {\bibfnamefont {H.}~\bibnamefont {Simpson}}, \bibinfo {author} {\bibfnamefont {S.}~\bibnamefont {Kouchekian}},\ and\ \bibinfo {author} {\bibfnamefont {S.}~\bibnamefont {Yakubovich}},\ }\bibfield  {title} {\bibinfo {title} {On the orthogonality of the macdonald's functions},\ }\href {https://doi.org/https://doi.org/10.1016/j.jmaa.2009.06.067} {\bibfield  {journal} {\bibinfo  {journal} {Journal of Mathematical Analysis and Applications}\ }\textbf {\bibinfo {volume} {360}},\ \bibinfo {pages} {380} (\bibinfo {year} {2009})}\BibitemShut {NoStop}%
\bibitem [{\citenamefont {Ishikawa}(2024)}]{ISHIKAWA2024169571}%
  \BibitemOpen
  \bibfield  {author} {\bibinfo {author} {\bibfnamefont {K.}~\bibnamefont {Ishikawa}},\ }\bibfield  {title} {\bibinfo {title} {Potential scatterings in the {L}2 space : (2) rigorous scattering probability of wave packets},\ }\href {https://doi.org/https://doi.org/10.1016/j.aop.2023.169571} {\bibfield  {journal} {\bibinfo  {journal} {Annals of Physics}\ }\textbf {\bibinfo {volume} {460}},\ \bibinfo {pages} {169571} (\bibinfo {year} {2024})}\BibitemShut {NoStop}%
\bibitem [{\citenamefont {Bojowald}\ \emph {et~al.}(2024)\citenamefont {Bojowald}, \citenamefont {Brizuela}, \citenamefont {Calizaya~Cabrera},\ and\ \citenamefont {Uria}}]{Bojowald:2023fas}%
  \BibitemOpen
  \bibfield  {author} {\bibinfo {author} {\bibfnamefont {M.}~\bibnamefont {Bojowald}}, \bibinfo {author} {\bibfnamefont {D.}~\bibnamefont {Brizuela}}, \bibinfo {author} {\bibfnamefont {P.}~\bibnamefont {Calizaya~Cabrera}},\ and\ \bibinfo {author} {\bibfnamefont {S.~F.}\ \bibnamefont {Uria}},\ }\bibfield  {title} {\bibinfo {title} {{Chaotic behavior of the Bianchi IX model under the influence of quantum effects}},\ }\href {https://doi.org/10.1103/PhysRevD.109.044038} {\bibfield  {journal} {\bibinfo  {journal} {Phys. Rev. D}\ }\textbf {\bibinfo {volume} {109}},\ \bibinfo {pages} {044038} (\bibinfo {year} {2024})},\ \Eprint {https://arxiv.org/abs/2307.00063} {arXiv:2307.00063 [gr-qc]} \BibitemShut {NoStop}%
\bibitem [{\citenamefont {Bojowald}\ \emph {et~al.}(2023)\citenamefont {Bojowald}, \citenamefont {Brizuela}, \citenamefont {Calizaya~Cabrera},\ and\ \citenamefont {Uria}}]{Bojowald:2023sjw}%
  \BibitemOpen
  \bibfield  {author} {\bibinfo {author} {\bibfnamefont {M.}~\bibnamefont {Bojowald}}, \bibinfo {author} {\bibfnamefont {D.}~\bibnamefont {Brizuela}}, \bibinfo {author} {\bibfnamefont {P.}~\bibnamefont {Calizaya~Cabrera}},\ and\ \bibinfo {author} {\bibfnamefont {S.~F.}\ \bibnamefont {Uria}},\ }\bibfield  {title} {\bibinfo {title} {{Reduction of primordial chaos by generic quantum effects}},\ }\href {https://doi.org/10.1103/PhysRevD.108.L061501} {\bibfield  {journal} {\bibinfo  {journal} {Phys. Rev. D}\ }\textbf {\bibinfo {volume} {108}},\ \bibinfo {pages} {L061501} (\bibinfo {year} {2023})},\ \Eprint {https://arxiv.org/abs/2307.13040} {arXiv:2307.13040 [gr-qc]} \BibitemShut {NoStop}%
\bibitem [{\citenamefont {Brizuela}\ and\ \citenamefont {Uria}(2022)}]{Brizuela:2022uun}%
  \BibitemOpen
  \bibfield  {author} {\bibinfo {author} {\bibfnamefont {D.}~\bibnamefont {Brizuela}}\ and\ \bibinfo {author} {\bibfnamefont {S.~F.}\ \bibnamefont {Uria}},\ }\bibfield  {title} {\bibinfo {title} {{Semiclassical study of the mixmaster model: The quantum Kasner map}},\ }\href {https://doi.org/10.1103/PhysRevD.106.064051} {\bibfield  {journal} {\bibinfo  {journal} {Phys. Rev. D}\ }\textbf {\bibinfo {volume} {106}},\ \bibinfo {pages} {064051} (\bibinfo {year} {2022})},\ \Eprint {https://arxiv.org/abs/2207.00566} {arXiv:2207.00566 [gr-qc]} \BibitemShut {NoStop}%
\bibitem [{\citenamefont {Uria}\ \emph {et~al.}(2023)\citenamefont {Uria}, \citenamefont {Brizuela},\ and\ \citenamefont {Alonso-Serrano}}]{Uria:2023uik}%
  \BibitemOpen
  \bibfield  {author} {\bibinfo {author} {\bibfnamefont {S.~F.}\ \bibnamefont {Uria}}, \bibinfo {author} {\bibfnamefont {D.}~\bibnamefont {Brizuela}},\ and\ \bibinfo {author} {\bibfnamefont {A.}~\bibnamefont {Alonso-Serrano}},\ }\bibfield  {title} {\bibinfo {title} {{Quantum corrections to the Bianchi II transition under local rotational invariance}},\ }in\ \href {https://doi.org/10.1142/9789811269776_0044} {\emph {\bibinfo {booktitle} {{16th Marcel Grossmann Meeting on~Recent Developments in Theoretical and Experimental General Relativity, Astrophysics and Relativistic Field Theories}}}}\ (\bibinfo {year} {2023})\BibitemShut {NoStop}%
\end{thebibliography}%

\end{document}